\documentclass[aps,prl,twocolumn]{revtex4}
\usepackage{graphicx}

\newcommand{\Eb}{\mbox{\bf E}}

\newcommand{\ub}{\mbox{\bf u}}

\begin{document}

\title{ Nonlinear electrokinetics at large applied voltages  }

\author{Martin Z. Bazant$^{1,2}$, Mustafa Sabri Kilic$^1$,
Brian D. Storey$^3$, and Armand Ajdari$^{1,4}$}
\affiliation{$^1$ Department of Mathematics, Massachusetts Institute
  of Technology, Cambridge, MA 02139 \\
$^2$  Institute for Soldier
Nanotechnologies, Massachusetts Institute of Technology, Cambridge, MA
02139\\
$^3$ Franklin W. Olin College of Engineering, Needham, MA 02492  \\
$^4$ UMR Gulliver ESPCI-CNRS 7083, 10 rue Vauquelin, F-75005
Paris, France}

\date{\today}

\begin{abstract}
  The classical theory of electrokinetic phenomena assumes a dilute
  solution of point-like ions in chemical equilibrium with a surface
  whose double-layer voltage is of order the thermal voltage, $k_BT/e
  = 25$ mV. In nonlinear ``induced-charge'' electrokinetic phenomena,
  such as AC electro-osmosis, several Volts $\approx 100 k_BT/e$ are
  applied to the double layer, so the theory breaks down and cannot
  explain many observed features. We argue that, under such a large
  voltage, counterions ``condense'' near the surface, even for dilute
  bulk solutions. Based on simple models, we predict that the
  double-layer capacitance decreases and the electro-osmotic mobility
  saturates at large voltages, due to steric repulsion and increased
  viscosity of the condensed layer, respectively. The former suffices
  to explain observed high frequency flow reversal in AC
  electro-osmosis; the latter leads to a salt concentration dependence
  of induced-charge flows comparable to experiments, although a
  complete theory is still lacking.
\end{abstract}

\maketitle

Electrically driven flows in ionic solutions are finding many new
applications in microfluidics~\cite{squires2005}. The theory of
electro-osmosis~\cite{lyklema_book_vol2} was developed for slip past a
surface in chemical {\it equilibrium}, whose double-layer voltage is
typically of order ${k_BT}/{e} = 25$ mV. However the discovery of AC
electro-osmosis (ACEO) at micro-electrodes
~\cite{ramos1999,ajdari2000} has shifted attention to a new regime,
where the {\it induced} double-layer voltage is $\approx 100\, k_BT/e$,
oscillating at frequencies up to 100 kHz, and nonuniform at the micron
scale. Related effects of induced-charge electro-osmosis
(ICEO)~\cite{iceo2004a,iceo2004b} also occur around colloidal
particles~\cite{murtsovkin1996} and polarizable
microstructures~\cite{levitan2005} (in AC or DC fields), and
asymmetric particles can move by induced-charge electrophoresis
(ICEP)~\cite{iceo2004a,squires2006}.

In all of these situations, low-voltage theories fail to predict
crucial experimental trends~\cite{olesen2006}, such as flow decay at
high salt concentration~\cite{studer2004,levitan_thesis,velev} and flow
reversal in asymmetric pumps at high voltage and high frequency~\cite{studer2004,chang2004,urbanski2006}. In this Letter, we
attribute these failures to the breakdown of the dilute solution
approximation at large applied voltages.  Based on very simple models,
we predict two general effects due to counterion crowding -- decay of
the double-layer capacitance and saturation of the electro-osmotic
mobility -- which begin to explain the experimental data.

{\it Experimental puzzles -} ICEO flows are rather complex, so many
simplifications have been made to arrive at an operational
model~\cite{murtsovkin1996,ramos1999,ajdari2000,iceo2004b,squires2006}.
For thin double layers, the first step is to solve Laplace's equation
(Ohm's Law) for the electrostatic potential in the conducting bulk,
$\nabla^2\phi=0$, with a capacitance-like  boundary condition
to close the ``RC'' circuit~\cite{bazant2004},
\begin{equation}
C_D \frac{d\Psi_D}{dt} = \sigma E_n,    \label{eq:RC}
\end{equation}
where the local diffuse-layer voltage $\Psi_D(\phi)$ responds to the
normal electric field $E_n=-\hat{n}\cdot \nabla \phi$; the bulk
conductivity $\sigma$ and diffuse-layer capacitance $C_D$ are usually
constants, although these assumptions can be
relaxed~\cite{olesen2006,chu2006}.  The second step is to solve for a
Stokes flow with the Helmholtz-Smoluchowski effective slip boundary condition,
\begin{equation}
  \ub_s = -b \, \Eb_t = -\frac{\varepsilon_b \zeta}{\eta_b}\, \Eb_t  \label{eq:HS}
\end{equation}
where $b(\Psi_D)$ is the electro-osmotic mobility, $\Eb_t$ the
tangential field, $\zeta = \Psi_D$ is the ``zeta potential'' at the
shear plane ($\zeta = \Psi_D$ in the simplest models),
and $\varepsilon_b$ and $\eta_b$ are the permittivity and
viscosity of the {\it bulk} solvent. While this model
based on dilute solution theory  can only be justified
for $\Psi_D \approx k_BT/e$
~\cite{gonzalez2000,iceo2004b,bazant2004},
it  describes many features of ICEO flows at moderate and large
voltages. Nevertheless, some crucial effects are still missing.

ICEO flows have a strong sensitivity to solution chemistry, which
is under-reported and unexplained. Recent experiments suggest a
universal logarithmic decay of the mobility, $b \propto
\log(c_c/c_0)$, with bulk concentration $c_0$ seen in KCl for ACEO
micropumps~\cite{studer2004}, in KCl and CaCo$_3$ for ICEO flows
around metal posts~\cite{levitan_thesis}, and in NaCl for ICEP of
colloids with partial metal coatings~\cite{velev}. In all cases, the
flow practically vanishes for $c_0 > c_c \approx 30$ mM, well below
the concentration of most biological fluids ($c_0\approx 0.3$ M).
ICEO flows~\cite{levitan_thesis} and AC-field induced interactions in
colloids~\cite{sides2001} are also sensitive to the particular ions, at a given
concentration.

The low-voltage model also fails to describe the reversal of ICEO flow
observed in some (but not all) experimental situations. Flow reversal
was first reported around metal particles in
water~\cite{gamayunov1992}, where the velocity agreed with the
theory~\cite{murtsovkin1996} only for micron-sized particles and
reversed for larger ones ($90-400 \mu$m). The transition occured when
several volts was applied across the particle and reversal was
attributed to Faradaic reactions~\cite{gamayunov1992}. Flow reversal
has also been observed at high voltage ($> 2$ V) and high frequency
(10-100 kHz) in ACEO pumping of dilute KCl~\cite{studer2004,chang2004}
and deionized water~\cite{urbanski2006} with $10 \mu$m scale electrode
arrays. This reversal was first attributed to Faradaic
reactions~\cite{chang2004}, but simulations with Butler-Volmer
reaction kinetics have failed to predict the observed
flow~\cite{olesen2006}.  With non-planar 3D
electrodes~\cite{bazant2006}, the low-voltage model also fails to
predict the double-peaked frequency spectrum, which accompanies flow
reversal in some geometries~\cite{urbanski2006}.

Although Faradaic reactions surely occur at large voltages, they are
dominant at low frequencies in ACEO simulations~\cite{olesen2006} and
experiments (when gas bubbles arise)~\cite{studer2004}. Dilute solution
theories also predict that
nonlinear effects dominate at low
frequencies: The differential capacitance of
the diffuse layer~\cite{lyklema_book_vol2},
\begin{equation}
  C_{D}(\Psi_D)=\frac{\varepsilon_b}{\lambda_{D}}\cosh\left(\frac{ze\Psi_{D}}{2k_BT}
  \right) \label{eq:cdpb}
\end{equation}
causes the RC charging time to grow exponentially with
voltage~\cite{olesen2006}, and salt adsorption and tangential
conduction by the diffuse layer are coupled to (much slower) bulk
diffusion~\cite{bazant2004,chu2006}. Strong flow reversal is
experimentally observed at much higher frequencies, and the
unexplained concentration dependence seems to be independent of
frequency.  We conclude that dilute solution theories do not properly
describe the dynamics of electrolytes at large voltages.

\begin{figure}
(a)\includegraphics[width=2.8in]{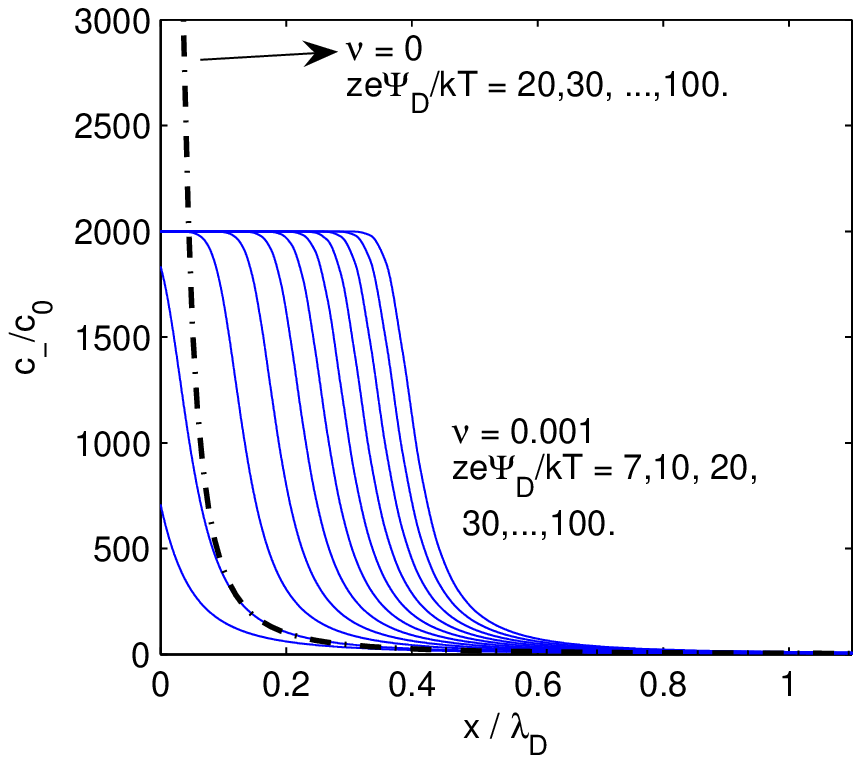}
(b)\includegraphics[width=2.8in]{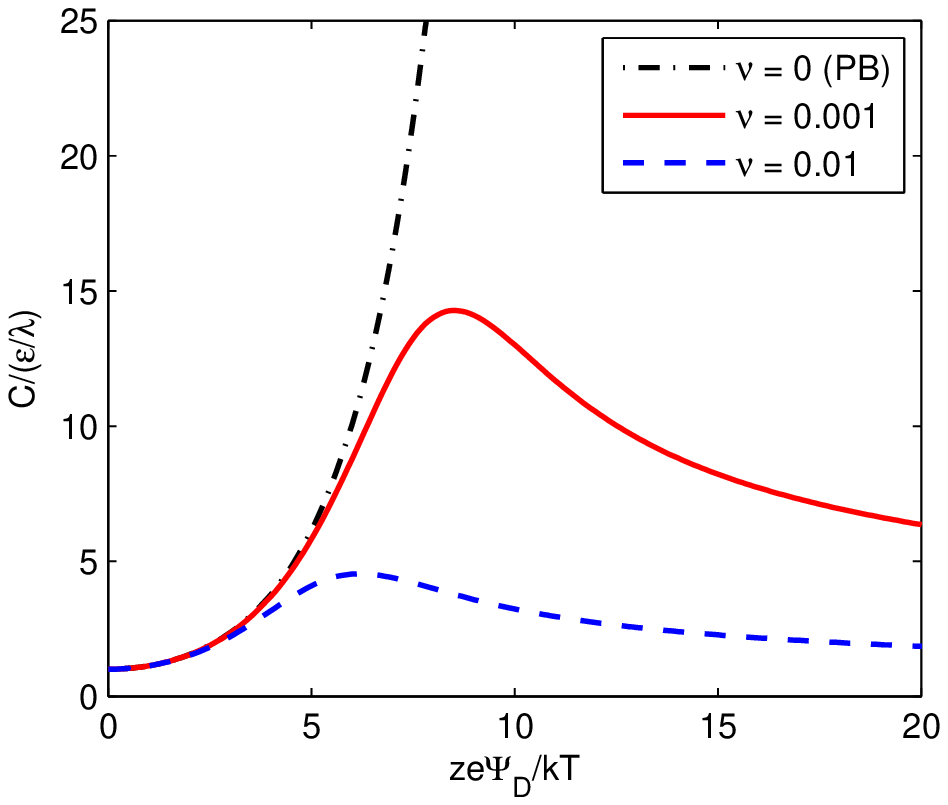}
\caption{\label{fig:C} (a) The equilibrium distribution of counterions
  near a surface in Poisson-Boltzmann (PB) and a modified (MPB) theory
  taking into account a minimum ion spacing $a$ for large applied
  voltages $ze\Psi_D/k_BT=20,30,\ldots, 100$, also $ze\Psi_D/k_BT=7,10$
  are shown for MPB to illustrate intermediate stages. (b) The voltage dependence
  of the differential capacitance $C_D$ of the diffuse layer from
  (\ref{eq:cdpb}) and (\ref{eq:cdnu}), where $\nu = 2a^3 c_0$ is the
  bulk volume fraction of ions.  }
\end{figure}

{\it Crowding effects -} All dilute solution theories, which describe
point-like ions in a mean-field approximation, break down when the
crowding of ions become significant, when steric effects and
correlations potentially become important.  If this can be translated
into a characteristic length scale $a$ for the distance between ions,
then the validity of the Poisson-Boltzmann (PB) is limited by a cutoff
concentration $c_{max}=a^{-3}$, reached at a fairly small voltage,
\begin{equation}
\Psi_c = - \frac{k_BT}{ze} \ln(a^3c_0) =   \frac{k_BT}{ze}
\ln\left(\frac{c_{max}}{c_0} \right).  \label{eq:psic}
\end{equation}
where $z$ is the valence and $c_0$ the bulk concentration of the
counterions. In a dilute solution of small ions, this
leads to cutoffs well below typical voltages for ICEO flows. For example,
even if only steric effects are taken into account, with e.g.
$a=3$ \AA (including a solvation shell), then
$\Psi_c \approx 0.33 V$ for $c_0=10^{-5}$ M and $z=1$.
To account for the obvious excess ions in PB theory, Stern
long ago postulated a static compact monolayer of solvated ions~\cite{bockris_book}.
This is also invoked in some models of ICEO flows, where
a constant capacitance is added to model the Stern layer and/or a
dielectric coating, which carries most of the voltage when the
diffuse-layer capacitance (\ref{eq:cdpb}) diverges.
However the voltage drops applied in ICEO make it unrealistic that a monolayer
could withstand most of the drop, and furthermore a dynamical model
is required for a condensed layer that is built and destroyed as the applied
field alternates.


A variety of ``modified Poisson-Boltzmann equations'' (MPB) have been
proposed to describe equilibrium ion profiles near a charged
wall~\cite{kilic2006a,biesheuvel2007}.  To capture ion crowding effects
across a wide range of voltages, we employ the simplest possible MPB
model of Bikerman~\cite{bikerman1942} and
others~\cite{kilic2006a,borukhov1997}, which is a continuum
approximation of the entropy of ions on a lattice of size $a$. As
shown in Fig.~\ref{fig:C}(a), when a large voltage is applied, the
counterion concentration exhibits a smooth transition from an outer PB
profile to a condensed layer at $c=c_{max}=a^{-3}$ near the surface.

Equation (\ref{eq:cdpb}) of dilute-solution theory predicts that $C_D$
diverges with $\Psi_D$, but with this model we predict the opposite
dependence ~\cite{kilic2006a},
\begin{equation}
C_{D}^{\nu }=
\frac{\frac{\varepsilon}{\lambda_{D}}\sinh(\frac{ze\Psi_{D}}{k_BT})}
{[1+2\nu\sinh^2\left(\frac{ze\Psi_{D}}{2k_BT}\right)]
\sqrt{\frac{2}{\nu}[1+2\nu\sinh^2\left(\frac{ze\Psi_{D}}{2k_BT}\right)]}}
\label{eq:cdnu}%
\end{equation}
where $\nu = 2a^3 c_0$ is the bulk volume fraction of ions.  As shown
in Fig.~\ref{fig:C}(b), the capacitance reaches a maximum near the
critical voltage $\Psi_c$ and then {\it decreases} at larger voltages
because the effective capacitor width grows due to steric effects.

\begin{figure}
(a) \includegraphics[width=3in]{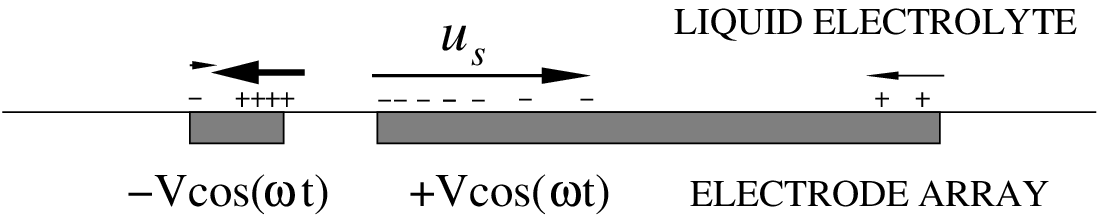} \\
\ \\
(b) \includegraphics[width=2.8in]{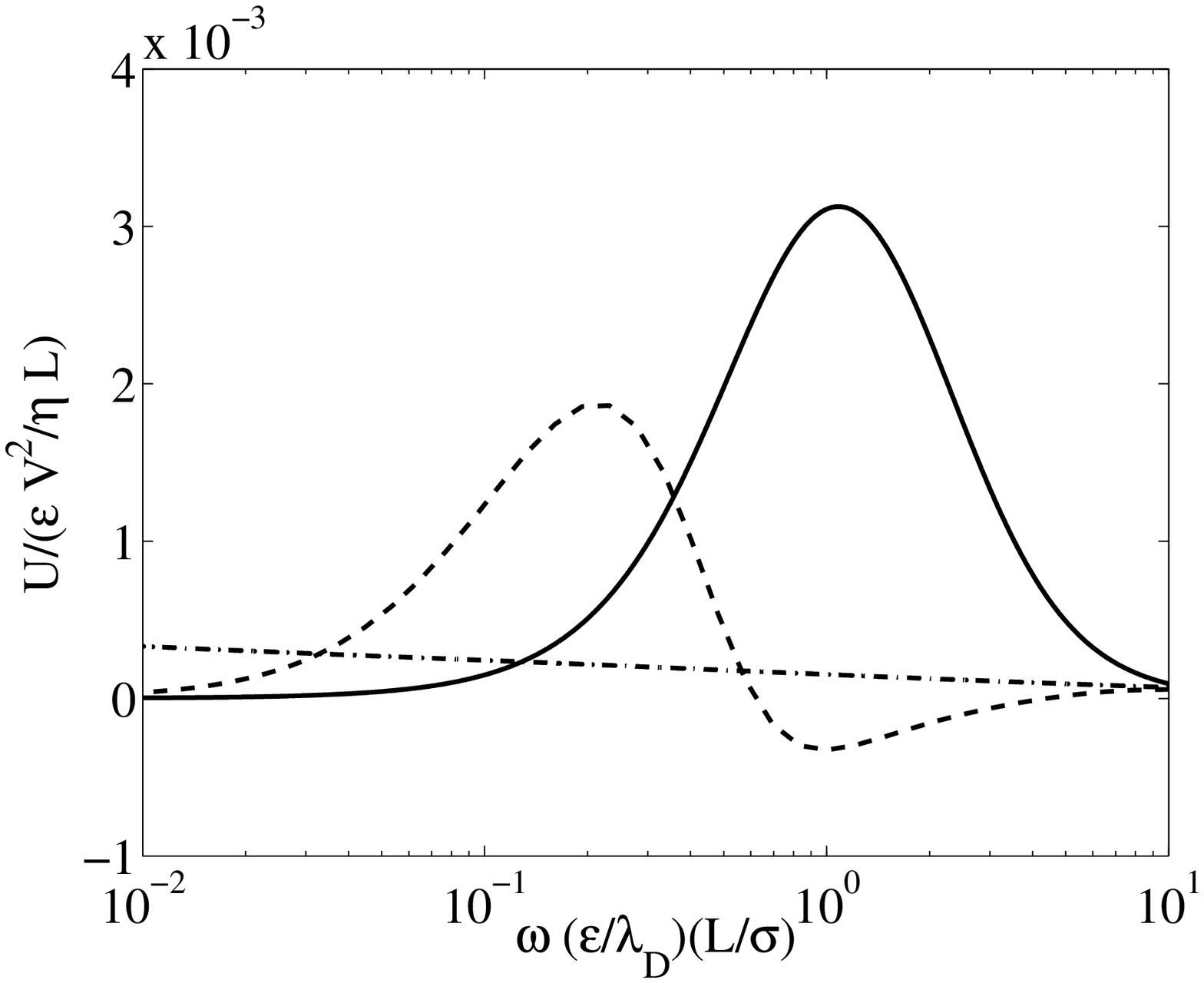}
\caption{\label{fig:rev} (a) Sketch of one period of the most studied
  ACEO pump geometry~\cite{brown2001}, indicating slip (arrows) and
  induced diffuse-layer charge, and (b) the (dimensionless) flow rate
  versus frequency for different models. In the low-voltage limit
  $V\ll k_BT/e=25$ mV, all models predict a single peak (solid
  line). For a typical experimental voltage, $V=100 k_BT/e = 2.5$ V,
  Poisson-Boltzmann theory breaks downs and its capacitance
  (\ref{eq:cdpb}) shifts the flow to very low frequency (dot-dashed
  line), but accounting for steric effects (\ref{eq:cdnu}) with $\nu = 0.001$
  (dashed line) reduces the shift and predicts high
  frequency flow reversal, similar to experiments~\cite{studer2004}.  }
\end{figure}

This decrease of diffuse-layer capacitance at large voltages is robust
to variations in the model and has major consequences for nonlinear
electrokinetics. For example, it provides a simple explanation for the
flow reversal seen in ACEO experiments, without invoking Faradaic
reactions.  As shown in Fig.~\ref{fig:rev}, numerical simulations of a
well studied planar pump geometry~\cite{brown2001,studer2004} with the
standard linear model~\cite{ramos1999,ajdari2000,ramos2003,bazant2006}
predict a single peak in flow rate versus frequency at all
voltages. If the nonlinear PB capacitance (\ref{eq:cdpb}) is
included~\cite{olesen2006}, then the peak is reduced and shifts to
much lower frequency (contrary to experiments), due to slower charging
dynamics at large voltage~\cite{bazant2004,chu2006}. The MPB
capacitance with steric effects (\ref{eq:cdnu}) reduces the peak shift
and introduces flow reversal similar to experiments, albeit with a
large value of $a = 0.0005 c_0^{-1}$, perhaps due to the
underprediction of liquid steric effects by the lattice
approximation~\cite{biesheuvel2007}.

The physical mechanism for flow reversal is simple: At low voltage,
the pumping direction is set by the larger electrode, but at large
voltages, since the more highly charged smaller electrode has its
``RC'' charging time reduced by steric effects, it ``wins'' in pumping
against the larger electrode at higher frequency. Perhaps a similar
effect is responsible for the double-peaked structure at high voltage
with non-planar stepped electrodes~\cite{urbanski2006}.


{\it Induced viscosity increase - } The strong decay of ICEO flow
with inceasing concentration suggests a dramatic increase in the
viscosity of a highly charged diffuse layer.  Classical continuum
theory provides a general formula for the electro-osmotic
mobility~\cite{lyklema_book_vol2},
\begin{equation}
b = \int_0^{\Psi_D} \frac{\varepsilon}{\eta} d\Psi = \frac{\varepsilon_b \overline{\zeta}}{\eta_b}
\end{equation}
as an integral over the potential difference $\Psi$ entering the
diffuse layer from the bulk. At large voltages, the effective zeta
potential $\overline{\zeta}$ (a measure of flow generated) should
be smaller than $\Psi_D$ as $\varepsilon$ decreases (due to
alignment of water dipoles) and $\eta$ increases (due to viscoelectric
effects) within the diffuse layer.

Focusing on the viscoelectric effect in water, Lyklema and
Overbeek~\cite{lyklema1961,lyklema1994} first derived a modified slip
formula by assuming $\eta \propto E^2$ in PB theory and predicted that
$b(\Psi_D)$ saturates at a constant value, which decays with
increasing $c_0$. The saturation, however, relies on the unphysical
divergence of the counterion concentration (and thus $E$) in PB
theory; with MPB steric effects included, it can be shown that the
mobility does not saturate, and the viscoelectric effect is not strong
enough to describe ICEO experiments.

\begin{figure}
\includegraphics[width=3in]{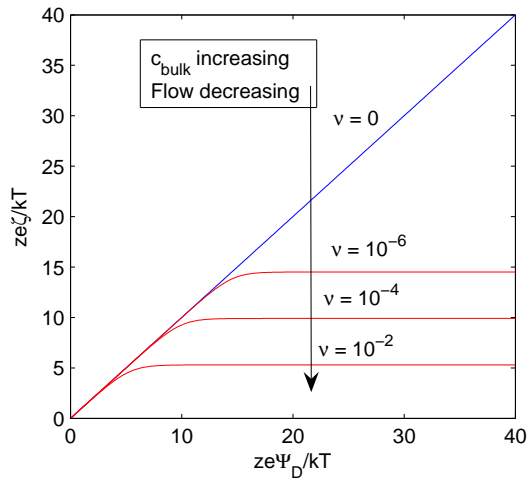}
\caption{ \label{fig:zeta} Effective zeta potential $\overline{\zeta}
  \equiv u_s\eta_b/E_t\varepsilon_b$ (slip velocity per tangential
  electric field) versus diffuse-layer voltage $\Psi_D$ at different bulk
  concentrations $c_0$ for the modified slip formula
  (\ref{eq:mslip}) postulating a viscoelectric effect related to
  crowding of ions in the diffuse layer at large voltages.
}
\end{figure}

In a very crude attempt, we adopt the MPB theory above and postulate
that $\varepsilon/\eta$ diverges as the counterion density (equivalent
to the charge density $\rho$) approaches the
steric limit $c_{max}$ as:
\begin{equation}
\frac{\varepsilon}{\eta} = \frac{\varepsilon_b}{\eta_b} \left( 1 -
  \frac{\rho}{c_{max}}\right) = \frac{\varepsilon_b(1 - a^3\rho)}{\eta_b}.
\end{equation}
This (arbitrary) choice leads to a simple formula for the effective zeta potential,
\begin{equation}
  \overline{\zeta} = \Psi_D - \mbox{sgn}(\Psi_D)\, \frac{k_BT}{ze} \,
  \log\left[ 1 + 4 a^3 c_0 \sinh^{2}\left( \frac{ze\Psi_D}{2k_BT}
    \right)\right]    \label{eq:mslip}
\end{equation}
which reduces to (\ref{eq:HS}) ($\overline{\zeta} \sim \Psi_D$) for
$\Psi_D \ll \Psi_c$ but saturates $\overline{\zeta}\sim \Psi_c$ for
$\Psi_D \gg \Psi_c$. From (\ref{eq:psic}), we recover the
experimentally observed scaling with concentration in the large
voltage limit, $u \propto \log(c_c/c_0)$, with $c_c = c_{max} =
a^{-3}$.  The mobility $b = \varepsilon_b \overline{\zeta}/\eta_b$
from (\ref{eq:mslip}) is also sensitive to the solution chemistry,
through $a$, $z$, and $c_0$, unlike the classical formula
(\ref{eq:HS}) valid at low voltages. The saturation of
$\overline{\zeta}$ also implies that the scaling of ICEO flows changes
from quadratic, $u \propto E^2$ or $V^2$, to linear, $u \propto |E|$
or $|V|$ at large voltages.

These predictions make the theory more realistic, but the
experimentally observed $c_c = 10$ mM implies a mean ion spacing
of $a = 4.4$ nm (roughly 40 atomic diameters) for the divergence of
$\varepsilon/\eta$.
How could we explain such a large value?  Perhaps at large voltages,
counterions condense into a sort of Wigner crystal, which resists
shear due to strong electrostatic correlations (in addition to
viscoelectric effects in the solvent). Indeed, the mean-field
approximation breaks down when ion spacings approach the Bjerrum
length, $l_B = (ze)^2/4\pi\varepsilon k_BT$, which is 7~ \AA ~ for bulk
water and monovalent ions ($z=1$). If $\varepsilon \approx 0.1
\varepsilon_b$ (as electrochemists infer for the Stern
layer~\cite{bockris_book}), then $a \approx l_B$ is possible, so
correlation effects on electro-osmotic flow (which to our knowledge
have never been studied) could be very significant at large voltages,
even in dilute bulk solutions.

In conclusion, we have argued that (at least) two new phenomena arise
in electro-osmosis at large induced voltages: (i) Crowding effects
decrease the differential capacitance (Fig. 1) which can explain high
frequency flow reversal in ACEO pumps (Fig. 2); (ii) viscosity
increase upon ion crowding saturates the mobility (Fig. 3), which
implies dependence on solution chemistry and flow decay with
increasing concentration. Although we believe these predictions are
robust we have not managed to combine our simplist models
(\ref{eq:cdnu}) and (\ref{eq:mslip}) into a complete theory. For
example, choosing $a \approx 1-4$ nm in (\ref{eq:mslip}) to fit the
critical concentration $c_c \approx 0.01-0.05 $ M tends to eliminate
flow reversal in Fig. 2b since the reduced mobility of small electrode
in Fig. 2a offsets its faster charging. Choosing a ``steric'' value $a
\approx 1-4$ ~\AA~ in (\ref{eq:cdnu}) shifts the flow to too low
frequency in dilute solutions (as in PB theory~\cite{olesen2006}) and
overestimates the concentration scale for ICEO flow suppression in
(\ref{eq:mslip}). More realistic MPB
models, which predict stronger steric effects~\cite{biesheuvel2007},
may improve the fit, but correlation effects on ICEO flow may also
need to be described.

Of course, our models are over-simplified, but remember that the
challenge is to describe ICEO flow over more than three decades of
diffuse-layer voltage from $k_BT/e = 25$ mV to $\approx 10$~V.  The
upper limit corresponds to a new regime for the theory of
electro-osmosis, where counterions are condensed near a highly charged
surface. Nanoscale experiments and atomic-level simulations will be
crucial to further develop the theory.

%
%

\

This research was supported in part by the U.S. Army through the
Institute for Soldier Nanotechnologies, under Contract DAAD-19-02-0002
with the U.S. Army Research Office. AA acknowledges the hospitality of
MIT and financial help from ANR grant Nanodrive.

\end{document}